\documentclass[10pt]{article}

\usepackage{amsmath}
\usepackage{amsfonts}
\usepackage{amssymb}
\usepackage{amsthm}
\usepackage{stmaryrd} 
\usepackage{braket}
\usepackage{empheq}
\usepackage{graphicx}
\usepackage{subfigure}
\usepackage{hyperref}
\usepackage{footmisc}
\usepackage{appendix}
\usepackage{epsf}
\usepackage{epsfig}
\usepackage{tikz,tikz-3dplot}
\usetikzlibrary{decorations.markings,decorations.pathmorphing}
\usepackage{caption}
\usepackage{pgfplots}
\usepackage[utf8]{inputenc}
\usepackage[english]{babel}
\usepackage[T1]{fontenc}
\usepackage{sectsty}
\subsubsectionfont{\normalfont\itshape}
\usepackage{fullpage}

\renewcommand{\bar}{\overline}
\renewcommand{\leq}{\leqslant}
\renewcommand{\geq}{\geqslant}

\newcommand{\triplenorm}[1]{|\mskip -2 mu|\mskip -2 mu|#1|\mskip -2mu |\mskip -2mu|}
\newcommand{\bbone}{{\text{\usefont{U}{bbold}{m}{n}\char49}}}

\DeclareMathOperator{\tr}{tr}

\theoremstyle{plain}
\newtheorem{theorem}{Theorem}[section]
\newtheorem{proposition}[theorem]{Proposition}

\newtheorem{lemma}[theorem]{Lemma}

\theoremstyle{definition}

\newtheorem{remark}[theorem]{Remark}

\title{Decoherence as a high-dimensional geometrical phenomenon}
\author{\textsc{Antoine Soulas} \\ Institut de recherche mathématique de Rennes (IRMAR) \\ Université de Rennes et CNRS,  France \\  \small{Email: antoine.soulas@univ-rennes1.fr}  \\  \small{ORCID: 0000-0003-0952-6060}}
\date{}

\begin{document}
\maketitle

\abstract{We develop a mathematical formalism that allows to study decoherence with a great level generality, so as to make it appear as a geometrical phenomenon between reservoirs of dimensions. It enables us to give quantitative estimates of the level of decoherence induced by a purely random environment on a system according to their respectives sizes, and to exhibit some links with entanglement entropy and the entanglement area-law.}


\section*{Acknowledgements}
I would like to gratefully thank my PhD supervisor Dimitri Petritis for the great freedom he grants me in my research, while being nevertheless always present to guide me. I also thank my friends Dmitry Chernyak and Matthieu Dolbeault for illuminating discussions.

\section*{Declarations}
\textbf{Funding:} no funding was received for conducting this study. \\
\textbf{Competing Interests:} the author has no competing interests to declare that are relevant to the content of this article. \\
\textbf{Ethics approval, Consent, Data, Materials and code availability, Authors’ contribution statements:} not applicable. 

\newpage
\section*{Introduction} \label{intro}

The theory of decoherence is arguably one of the greatest advances in fundamental physics of the past forty years. Without adding anything new to the quantum mechanical framework, and considering that the Schrödinger equation is universally valid, it explains why quantum interferences virtually disappear at macroscopic scales. Since the pioneering papers \cite{zeh1970interpretation} \cite{zurek1981pointer}, a wide variety of models have been designed to understand decoherence in different specific contexts (see the review \cite{zurek2003decoherence} or \cite{joos1996decoherence} and the numerous references therein). In this paper, we would like to embrace a more general point of view and understand the mathematical reason why decoherence is so ubiquitous between quantum mechanical systems. 

We start by introducing general quantities and notations to present as concisely as possible the idea underlying the theory of decoherence (\S\ref{basics}). We then build two simple but very general models to reveal the mathematical mechanisms that make decoherence so universal, thereby justifying why quantum interferences disappear due to the Schrödinger dynamics only (\S\ref{first model} and \S\ref{second model}). We recover in \ref{pairs} and \ref{feels} the well-known typical decay of the non-diagonal terms of the density matrix in $n^{-\frac{1}{2}}$, with $n$ the dimension of the Hilbert space describing the environment. The most important result is Theorem \ref{estimate eta}, proved in \S\ref{eta}, giving estimates for the level of decoherence induced by a random environment on a system of given sizes. We conclude in \S\ref{consequences} that even very small environments (of typical size at least $N_{\mathcal{E}} = \ln(N_{\mathcal{S}})$ with $N_{\mathcal{S}}$ the size of the system) suffice, under assumptions discussed in \S\ref{hypotheses}. We also give a general formula estimating the level of classicality of a quantum system in terms of the entropy of entanglement with its environment (\S\ref{entropy}, proved in the annex \ref{annex}), and propose alternative ways of quantifying decoherence in \S\ref{alternative}. 

\section{The basics of decoherence} \label{basics}

The theory of decoherence sheds light on the reason why quantum interferences disappear when a system gets entangled with a macroscopic one, for example an electron in a double-slit experiment that doesn't interfere anymore when entangled with a detector. According to Di Biagio and Rovelli \cite{di2021stable}, the deep difference between classical and quantum is the way probabilities behave: all classical phenomena satisfy the total probability formula 
\[ \mathbb{P}(B=y) =  \sum_{x \in \mathrm{Im}(A)} \mathbb{P}(A=x) \mathbb{P}(B=y \mid A=x), \]
relying on the fact that, \textit{even though the actual value of the variable $A$ is not known, one can still assume that it has a definite value among the possible ones}. This, however, is not correct for quantum systems. It is well-known that the diagonal elements of the density matrix account for the classical behavior of a system (they correspond to the terms of the total probability formula) while the non-diagonal terms are the additional interference terms. As a reminder, this is because the probability to obtain an outcome $x$ is: \[ \tr(\rho \ket{x}\bra{x}) = \sum_{i,j=1}^n \rho_{ij} \braket{j \vert x}  \braket{x \vert i} = \sum_{i=1}^n \underbrace{\rho_{ii}  \lvert \braket{x \vert i} \rvert^2}_{\mathbb{P}(i) \mathbb{P}(x \mid  i)}  +  \sum_{1\leq i < j \leq n}  \underbrace{2 \mathrm{Re}( \rho_{ij}  \braket{j \vert x} \braket{x \vert i })}_{\text{interferences}}. \]

Here is the typical situation encountered in decoherence studies. Consider a system $\mathcal{S}$, described by a Hilbert space $\mathcal{H}_{\mathcal{S}}$ of dimension $d$, that interacts with an environment $\mathcal{E}$ described by a space $\mathcal{H}_{\mathcal{E}}$ of dimension $n$, and let $\mathcal{B} = (\ket{i})_{1\leq i \leq d}$ be an orthonormal basis of $\mathcal{H}_{\mathcal{S}}$. In the sequel, we will say that each $\ket{i}$ corresponds to a \textbf{possible history} of the system in this basis (this expression will be given its full meaning in a future article dedicated to the measurement problem). Let's also assume that $\mathcal{B}$ is a \textbf{conserved basis} during the interaction with $\mathcal{E}$. When $\mathcal{E}$ is a measurement apparatus for the observable $A$, the eigenbasis of $\hat{A}$ is clearly a conserved basis; in general, the eigenbasis of any observable such that $\hat{A} \otimes \bbone$ commutes with the interaction Hamiltonian is suitable (but the existence of such an observable is not guaranteed, unless $\hat{H}_{int}$ takes the form $\sum_i \hat{\Pi}^{\mathcal{S}}_i \otimes \hat{H}^{\mathcal{E}}_{i}$, where $(\hat{\Pi}^{\mathcal{S}}_i)_{1 \leq i \leq d}$ is a family of commuting orthogonal projectors).

We further suppose that $\mathcal{S}$ and $\mathcal{E}$ are initially non entangled, allowing us to write $\ket{\Psi} = \left( \sum_{i=1}^d c_i \ket{i} \right) \otimes \ket{\mathcal{E}_0}$ as the initial state before interaction. After a time $t$, due to its Schrödinger evolution in the conserved basis, the total state becomes $\ket{\Psi(t)} = \sum_{i=1}^d c_i \ket{i} \otimes \ket{\mathcal{E}_i(t)}$ for some unit vectors $(\ket{\mathcal{E}_i(t)})_{1\leq i \leq d}$. Define $\eta(t) \equiv \displaystyle \max_{i \neq j} \; \lvert \braket{\mathcal{E}_i(t) \vert \mathcal{E}_j(t)} \rvert$. If $(\ket{e_k})_{1\leq k \leq n}$ denotes an orthonormal basis of $\mathcal{H}_{\mathcal{E}}$, the state of $\mathcal{S}$, obtained by tracing out the environment, is: 
\begin{align*}
\rho_{\mathcal{S}}(t) &= \tr_{\mathcal{E}} \ket{\Psi(t)} \bra{\Psi(t)} \\
&= \sum_{k=1}^n \left( \sum_{i=1}^d \lvert c_i \rvert^2  \lvert \braket{e_k \vert \mathcal{E}_i (t)} \rvert^2 \ket{i}\bra{i} + \sum_{1 \leq i \neq j \leq d} c_i \bar{c_j} \braket{e_k \vert  \mathcal{E}_i (t)} \braket{\mathcal{E}_j (t) \vert e_k} \ket{i}\bra{j} \right) \\
&=   \sum_{i=1}^d \lvert c_i \rvert^2  \underbrace{ \sum_{k=1}^n \lvert \braket{e_k \vert \mathcal{E}_i (t)} \rvert^2 }_{= 1} \; \ket{i}\bra{i} + \sum_{1 \leq i \neq j \leq d} c_i \bar{c_j} \bra{\mathcal{E}_j (t)} \Big( \underbrace{ \sum_{k=1}^n \ket{e_k} \bra{e_k} }_{= \bbone} \Big) \ket{\mathcal{E}_i (t)} \; \ket{i}\bra{j} \\
&= \sum_{i=1}^d \lvert c_i \rvert^2 \ket{i}\bra{i} +  \sum_{1 \leq i \neq j \leq d} c_i \bar{c_j}  \braket{\mathcal{E}_j(t) \vert \mathcal{E}_i(t)} \ket{i}\bra{j} \\
&\equiv \rho_{\mathcal{S}}^{(d)} + \rho_{\mathcal{S}}^{(q)}(t),
\end{align*}
where $\rho_{\mathcal{S}}^{(d)}$ stands for the (time independent) diagonal part of $\rho_{\mathcal{S}}(t)$ (which corresponds to the total probability formula), and $\rho_{\mathcal{S}}^{(q)}(t)$ for the remaining non diagonal terms responsible for the interferences between the possible histories. It is not difficult to show (see the Annex \ref{annex}) that $\triplenorm{ \rho_{\mathcal{S}}^{(q)}(t)} \leq \eta(t)$, where $\triplenorm{M}$ stands for the usual operator norm on matrices, \textit{i.e.} $\triplenorm{M} = \sup_{\lVert \ket{\Psi} \rVert =1} \; \lVert M \ket{\Psi} \rVert$. Therefore $\eta$ \textit{measures how close the system is from being classical} because, as shown in \ref{annex}, we have for all subspaces $F \subset \mathcal{H}_{\mathcal{S}}$ (recall that, in the quantum formalism, probabilistic events correspond to subspaces): 

\begin{equation}
\lvert \underbrace{\tr(\rho_{\mathcal{S}}(t) \Pi_F)}_{\text{quantum probability}} -  \underbrace{\tr(\rho_{\mathcal{S}}^{(d)} \Pi_F)}_{\text{classical probability}} \rvert \leq \dim(F) \; \eta(t).   \label{classicality}
\end{equation}
In other words, $\eta(t)$ estimates how decohered the system is. Notice well that it is only during an interaction between $\mathcal{S}$ and $\mathcal{E}$ that decoherence can occur; any future internal evolution $U$ of $\mathcal{E}$ lets $\eta$ unchanged since $\braket{U \mathcal{E}_j \vert U \mathcal{E}_i} = \braket{\mathcal{E}_j \vert \mathcal{E}_i}$. Also, a more precise definition for $\eta$ could be $\displaystyle \max_{\substack{i \neq j \\ c_i, c_j \neq 0}} \; \lvert \braket{\mathcal{E}_i(t) \vert \mathcal{E}_j(t)} \rvert$ with, by convention, $\eta = 0$ when only one $c_i$ is non zero, so that $\rho_{\mathcal{S}}$ being diagonal in a basis becomes equivalent to $\eta=0$ in this basis. This way $\eta$ really quantifies the interferences between \textit{possible} histories (of non zero probability). This is not true with the definition above, as is clear for example for the trivial interaction $\ket{\Psi(t)} = \sum_{i=1}^d c_i \ket{i} \otimes \ket{\mathcal{E}_0}$: here $\rho_{\mathcal{S}}$ is diagonal (\textit{i.e.} no interferences) in any orthonormal basis containing the vector $\sum_{i=1}^d c_i \ket{i}$, but the simpler definition yields $\eta = 1$ in any basis.

The aim of the theory of decoherence is to explain why $\eta(t)$ rapidly goes to zero when $n$ is large, so that the state of the system almost immediately\footnote{It is actually very important that the decoherence process (in particular a measurement) is not instantaneous. Otherwise, it would be impossible to explain why an unstable nucleus continuously measured by a Geiger counter is not frozen due to the quantum Zeno effect. See the wonderful model of \cite[\S8.3 and \S8.4]{fonda1978decay} that quantifies the effect of continuous measurement on the decay rate.} evolves from $\rho_{\mathcal{S}}$ to $\rho_{\mathcal{S}}^{(d)}$ in the conserved basis. As recalled in the introduction, a lot of different models already explain this phenomenon in specific contexts. In this paper, we shall build two (excessively) simple but quite universal models that highlight the fundamental reason why $\eta(t) \rightarrow 0$ so quickly, and that will allow us to determine the typical size of an environment needed to entail proper decoherence on a system.

\section{First model: purely random environment} \label{first model}

When no particular assumption is made to specify the type of environment under study, the only reasonable behavior to assume for $\ket{\mathcal{E}_i(t)}$ is that of a Brownian motion on the sphere $\mathbb{S}^{n} = \{ \ket{\Psi} \in \mathcal{H}_{\mathcal{E}} \mid \lVert \ket{\Psi} \rVert = 1 \} \subset \mathcal{H}_{\mathcal{E}} \simeq \mathbb{C}^n \simeq \mathbb{R}^{2n}$. It boils down to representing the environment as a purely random system with no preferred direction of evolution. This choice will be discussed in \S\ref{hypotheses}. Another bold assumption would be the independence of the $(\ket{\mathcal{E}_i(t)})_{1\leq i \leq d}$; we will dare to make this assumption anyway. 

\subsection{Convergence to the uniform measure} 

We will first show that the probabilistic law of each $\ket{\mathcal{E}_i(t)}$ converges exponentially fast to the uniform probability measure on $\mathbb{S}^{n}$. To make things precise, endow $\mathbb{S}^{n}$ with its Borel $\sigma$-algebra $\mathcal{B}$ and with the canonical Riemannian metric $g$, which induces the uniform measure $\mu$ that we suppose normalized to a probability measure. Let $\nu_t$ be the law of the random variable $\ket{\mathcal{E}_i(t)}$, that is $ \nu_t(B) = \mathbb{P}\big(\ket{\mathcal{E}_i(t)} \in B \big)$ for all $B \in \mathcal{B}$. Denote $\Delta f = \frac{1}{\sqrt{g}} \partial_i (\sqrt{g} g^{ij} \partial_j f)$ the Laplacian operator on $\mathcal{C}^\infty(\mathbb{S}^{n})$ which can be extended to $L^2(\mathbb{S}^{n})$, the completion of $\mathcal{C}^\infty(\mathbb{S}^{n})$ for the scalar product $(f,h) = \int_{\mathbb{S}^{n}} f(x) h(x) \mathrm{d}\mu$. The Hille-Yosida theory allows to define the Brownian motion on the sphere as the Markov semigroup of stochastic kernels generated by $\Delta$. In particular, this implies that if $p_t$ is the density after a time $t$, \textit{i.e.} $\nu_t(\mathrm{d}x) = p_t(x) \mu(\mathrm{d}x)$, then $p_t = e^{t \Delta}p_0$. Of course, the law $\nu_0$ of the deterministic variable $\ket{\mathcal{E}_i(0)} = \ket{\mathcal{E}_0}$ corresponds to a Dirac distribution, which is not strictly speaking in $L^2(\mathbb{S}^{n})$, but we can rather consider it as given by a sharply peaked density (with respect to $\mu$) $p_0 \in L^2(\mathbb{S}^{n})$. Finally, recall that the total variation norm of a measure defined on $\mathcal{B}$ is given by $\lVert \sigma \rVert_{TV} =  \underset{B \in \mathcal{B} } \sup \lvert \sigma(B) \rvert$.

\begin{proposition}
We have $\lVert \nu_t - \mu \rVert_{TV} \underset{t \rightarrow +\infty} \longrightarrow 0$ exponentially fast. Moreover, if $T(\mathbb{S}^{n}) = \inf \{ t>0 \mid \lVert \nu_t - \mu \rVert_{TV} \leq \frac{1}{e} \}$ denotes the characteristic time to equilibrium for the brownian diffusion on $\mathbb{S}^{n}$, then $T(\mathbb{S}^{n}) \underset{n \rightarrow +\infty} \sim \frac{\ln(2n)}{4n}$.
\end{proposition}

\begin{proof}
See \cite{saloff1994precise} for a precise proof of this proposition. The overall idea is to decompose the density of the measure $\nu_t$ in an eigenbasis of the Laplacian, so that the Brownian motion (which is generated by $\Delta$) will exponentially kill all modes but the one associated with the eigenvalue 0, that is the constant one. The estimate of $T(\mathbb{S}^{n})$ is then obtained by examining how fast each mode (multiplied by its multiplicity) is killed. Interestingly enough, the convergence is faster as $n$ increases since $T(\mathbb{S}^{n}) \underset{n \rightarrow \infty} \longrightarrow 0$.

\end{proof}

\begin{remark}
Physically speaking, there is no reason for the $(\ket{\mathcal{E}_i(t)})_{1\leq i \leq d}$ to follow the canonical ‘unit speed’ Brownian motion used in the proof. Coupling constants $g_i$ should be introduced to replace $\Delta$ by $g_i \Delta$ for the diffusion of $\ket{\mathcal{E}_i(t)}$. But thanks to the exponential convergence — all the more efficient that the dimension $n$ is high, which will certainly be the case for macroscopic environments — the conclusion remains unchanged: the $(\ket{\mathcal{E}_i(t)})_{1\leq i \leq d}$ almost immediately follow the uniform law.
\end{remark}

\subsection{Most vectors are almost orthogonal} \label{pairs}

Consequently, we are now interested in the behavior of the scalar products between random vectors uniformly distributed on the complex $n$-sphere $\mathbb{S}^{n}$. The first thing to understand is that, \textit{in high dimension, most pairs of unit vectors are almost orthogonal}.

\begin{proposition} \label{variance S}
Denote by $S = \braket{ \mathcal{E}_1 \vert \mathcal{E}_2} \in \mathbb{C}$ the random variable where $\ket{\mathcal{E}_1}$ and $\ket{\mathcal{E}_2}$ are two independent uniform random variables on $\mathbb{S}^{n}$. Then $\mathbb{E}(S) = 0$ and $\mathbb{V}(S) = \mathbb{E}(\lvert S \rvert^2) = \frac{1}{n}$.
\end{proposition}

\begin{proof}
Clearly, $\ket{\mathcal{E}_1}$ and $-\ket{\mathcal{E}_1}$ have the same law, hence $\mathbb{E}(S) = \mathbb{E}(-S) = 0$. What about its variance? One can rotate the sphere to impose for example $\ket{\mathcal{E}_1} = (1,0, \dots , 0)$, and by independence $\ket{\mathcal{E}_2}$ still follows a uniform law. Such a uniform law can be achieved by generating $2n$ independent normal random variables $(X_i)_{1 \leq i \leq 2n }$ following $\mathcal{N}(0,1)$, and by considering the random vector $\ket{\mathcal{E}_2} = \left( \frac{X_1+ iX_2}{\sqrt{X_1^2 + \dots + X_{2n}^2 }} , \dots , \frac{X_{2n-1} + i X_{2n}}{ \sqrt{X_1^2 + \dots + X_{2n}^2 }} \right)$. Indeed, for any continuous function $f : \mathbb{S}^{n} \rightarrow \mathbb{R}$ (with $\mathrm{d}\sigma^n$ denoting the measure induced by Lebesgue's on $\mathbb{S}^{n}$): 
\begin{align*}
\mathbb{E}[f(\ket{\mathcal{E}_2})]
&= \frac{1}{(2\pi)^n} \int_{\mathbb{R}^{2n}} f \left( \frac{x_1+i x_2}{\sqrt{x_1^2 + \dots + x_{2n}^2 }}, \dots , \frac{x_{2n-1}+i x_{2n}}{\sqrt{x_1^2 + \dots + x_{2n}^2 }} \right) e^{-(x_1^2 + \dots + x_{2n}^2)/2} \mathrm{d}x_1 \dots  \mathrm{d}x_{2n} \\
&= \frac{1}{(2\pi)^n} \int_0^\infty \left[\int_{\mathbb{S}^{n}} f(u) \mathrm{d}\sigma^n(u) \right]e^{-\frac{r^2}{2}}r^{2n-1} \mathrm{d}r \\
&= \omega_n \int_{\mathbb{S}^{n}} f(u) \mathrm{d}\sigma^n(u),
\end{align*}
which means that $\ket{\mathcal{E}_2}$ defined this way follows indeed the uniform law. 

In these notations, $\lvert S \rvert^2 =\frac{X_1^2 + X_2^2}{X_1^2 + \dots + X_{2n}^2 }$. Since each $X_i^2$ follows a $\chi^2$ law, it is then a classical lemma to show that $\lvert S \rvert^2$ follows a $\beta_{1,n-1}$ distribution, whose mean equals $\frac{1}{n}$. For a more elementary argument, note that, up to relabelling the variables, we have $\forall k \in \llbracket 1,n \rrbracket, \; \mathbb{E} \left(\frac{X_1^2 + X_2^2}{X_1^2 + \dots + X_{2n}^2 } \right) =  \mathbb{E}\left(\frac{X_{2k-1}^2 + X_{2k}^2}{X_1^2 + \dots + X_{2n}^2} \right)$ and so:

\[ \mathbb{V}(S) = \mathbb{E} \left(\frac{X_1^2 + X_2^2}{X_1^2 + \dots + X_{2n}^2 } \right) = \frac{1}{n} \sum_{k=1}^n  \mathbb{E}\left(\frac{X_{2k-1}^2 + X_{2k}^2}{X_1^2 + \dots + X_{2n}^2} \right) =  \frac{1}{n}  \mathbb{E}(1) = \frac{1}{n}. \]
Alternatively, had we worked on the real sphere $\subset \mathbb{R}^{2n}$ endowed with the real scalar product, the variance would have been $\frac{1}{2n}$. This highlights the fact that the real and complex spheres are indeed isomorphic as topological or differential manifolds, but not as Riemannian manifolds.

The same result would have been recovered if, instead of picking randomly a pair of vectors, we had chosen uniformly the unitary evolution operators $(U^{(i)}(t))_{1\leq i \leq d}$ such that $\ket{\mathcal{E}_i(t)} = U^{(i)}(t) \ket{\mathcal{E}_0}$, resulting from the interaction Hamiltonian. Again, if no direction of evolution is preferred, it is reasonable to consider the law of each $U^{(i)}(t)$ to be given by the Haar measure $\mathrm{d}U$ on the unitary group $\mathcal{U}_n$. If moreover they are independent, then $U^{(i)}(t)^\dagger U^{(j)}(t)$ also follows the Haar measure for all $i,j$ so that, using \cite[(112)]{spengler2012composite}: 

\[ \mathbb{V} \big( \braket{\mathcal{E}_i(t) \vert \mathcal{E}_j(t)} \big) = \int_{\mathcal{U}_n} \lvert \braket{\mathcal{E}_0 \vert U \mathcal{E}_0} \rvert^2 dU = \prod_{i=2}^n \frac{i-1}{i} = \frac{1}{n}. \]
\end{proof}

Therefore, $\lvert \braket{\mathcal{E}_i(t) \vert \mathcal{E}_j(t)} \rvert$ is, after a very short time, of order $\sqrt{\mathbb{V}(S)} = \frac{1}{\sqrt{\dim(\mathcal{H}_\mathcal{E})}}$, which is a well-known estimate already obtained by Zurek in \cite{zurek1982environment}. When $d=2$, if $\mathcal{E}$ is composed of $N_{\mathcal{E}}$ particles and each of them is described by a $p$-dimensional Hilbert space, then very rapidly: 
\begin{equation} \eta \sim p^{-N_{\mathcal{E}}/2} \label{small_eta} \end{equation} 
which is virtually zero for macroscopic environments, therefore decoherence is guaranteed. Of course, this is not true anymore if $d$ is large, because there will be so many pairs that some of them will inevitably become non-negligible, and so will $\eta$. We would like to determine a condition between $n$ and $d$ under which proper decoherence is to be expected. In other words, what is the minimal size of an environment needed to decohere a given system? 

\begin{remark}
In fact, the assumption of the existence of a conserved basis made in \S\ref{basics} is practical but not really necessary. Indeed, for any basis $\mathcal{B} = (\ket{\varphi_i})_{1\leq i \leq d}$ of $\mathcal{H}_{\mathcal{S}}$, one can always decompose $\ket{\Psi(t)} = \sum_{i=1}^d c_i (t) \ket{\varphi_i} \otimes \ket{\mathcal{E}_i(t)}_{\mathcal{B}}$ for some unit vectors $(\ket{\mathcal{E}_i(t)}_{\mathcal{B}})_{1\leq i \leq d}$, where now the coefficients $c_i$ possibly depend on time. The brownian character and the independence of the $(\ket{\mathcal{E}_i(t)}_{\mathcal{B}})_{1\leq i \leq d}$ still suffices to lead to strong decoherence in $\mathcal{B}$. 

Seeking to characterize the basis in which decoherence mainly happens, or the so-called ‘pointer basis’, is an important and challenging aspect of the theory of decoherence. Several criteria have been proposed \cite{zurek1982environment} \cite{zurek2000decoherence}, but what precedes hints at a novel one (although physically very impractical): the pointer basis could be defined as the basis in which the $(\ket{\mathcal{E}_i(t)}_{\mathcal{B}})_{1\leq i \leq d}$ are independent, \textit{i.e.} the eigenbasis of their covariance matrix.
\end{remark}

\subsection{Direct study of $\eta$} \label{eta}

To answer this question, we should be more precise and consider directly the random variable $\eta_{n,d} =  \displaystyle \max_{i \neq j} \; \lvert \braket{\mathcal{E}_i \vert \mathcal{E}_j} \rvert$ where the $(\ket{\mathcal{E}_i)}_{1\leq i \leq d}$ are $d$ random vectors uniformly distributed on the complex $n$-sphere $\mathbb{S}^{n}$. In the following, we fix $\varepsilon \in \left]0,1 \right[$ as well as a threshold $s \in [0,1[$ close to $1$, and define $d_{max}^{\varepsilon, s}(n) = \min \{ d \in \mathbb{N} \mid \mathbb{P}(\eta_{n,d} \geq \varepsilon) \geq s \}$, so that if $d_{max}^{\varepsilon, s}(n)$ points or more are placed randomly on $\mathbb{S}^{n}$, it is very likely (with probability $\geq s$) that at least one of the scalar products will be greater than $\varepsilon$. 

\begin{theorem} \label{estimate eta}
The following asymptotic estimates hold:
\begin{enumerate}
\item $\displaystyle d_{max}^{\varepsilon, s}(n)  \underset{n \rightarrow \infty}{\sim} \sqrt{-2 \ln(1-s)} \left( \frac{1}{1-\varepsilon^2} \right)^{\frac{n-1}{2}}$
\item $\boxed{ \eta_{n,d} \underset{n \text{ or } d \rightarrow \infty}{\overset{\mathbb{P}}{ \xrightarrow{\hspace*{0.8cm}} }} \sqrt{1-d^{-\frac{2}{n}}} }$.
\end{enumerate}
\end{theorem}

To derive these formulas, we first need the following geometrical lemma. 

\begin{lemma} \label{lemma} 
Let $A_n = \lvert \mathbb{S}^{n} \rvert$ be the area of the complex $n$-sphere for $\mathrm{d}\sigma^n$ (induced by Lebesgue's measure), $C_{n}^{\varepsilon}(x) = \{ u \in \mathbb{S}^{n} \mid \lvert \braket{u \vert x } \rvert \geq \varepsilon \}$ the ‘spherical cap’\footnote{We use the quotation marks because, on $\mathbb{S}^{n}$ equipped with its complex scalar product, this set doesn't look like a cap as it does in the real case. QM is nothing but a geometrical way of calculating probabilities (in which the total probability formula is not true, so that it looks like all possible histories interfere), but the geometry in use is quite different from the intuitive one given by the familiar real scalar product. It is noteworthy to remark that the universe, through its quantum statistics, obeys very precisely the geometry of the \textit{complex} scalar product, and more generally the geometry induced by its canonical extension on tensor products of Hilbert spaces.} centered in $x$ of parameter $\varepsilon$, and $A_n^{\varepsilon} = \lvert C_n^{\varepsilon} \rvert$ the area of any spherical cap of parameter $\varepsilon$. Then for all $n\geq1$:
\[ \frac{A_n^{\varepsilon}}{A_n} = (1-\varepsilon^2)^{n-1}. \]
\end{lemma}

\begin{proof}[Proof of Lemma]
This result can be directly obtained from the fact that, as noticed in the proof of Proposition \ref{variance S}, $\lvert \braket{ \mathcal{E}_1 \vert \mathcal{E}_2} \rvert^2$ follows a $\beta_{1,n-1}$ distribution when $\ket{\mathcal{E}_1}$ and $\ket{\mathcal{E}_2}$ are chosen uniformly and independently on $\mathbb{S}^{n}$. We can then write:
\begin{align*}
\frac{A_n^{\varepsilon}}{A_n} =  \mathbb{P}(\lvert \braket{u \vert x } \rvert^2 \geq \varepsilon^2) = \int_{\varepsilon^2}^1 \frac{\Gamma(n)}{\Gamma(n-1)} (1-x)^{n-2} \mathrm{d}x = \left[ (1-x)^{n-1} \right]_{\varepsilon^2}^{1} = (1-\varepsilon^2)^{n-1}.
\end{align*}
A more ‘physicist-friendly’ proof can also be given, based on an appropriate choice of coordinates on the $n$-sphere. Recall that $\mathbb{S}^{n} \subset \mathbb{C}^n \simeq \mathbb{R}^{2n}$ can be seen as a real manifold of dimension $2n-1$. Consider the set of coordinates $(r,\theta, \varphi_1, \dots, \varphi_{2n-3})$ on $\mathbb{S}^{n}$ defined by the chart 

\[  \begin{array}{cccl} F :  & [0,1] \times [0,2\pi[ \times [0, \pi]^{2n-4} \times [0,2\pi[  &  \longrightarrow & \mathbb{S}^{n}  \\
                                           & (r, \theta, \varphi_1, \dots, \varphi_{2n-3}) & \longmapsto  & (x_1+ix_2, \dots, x_{2n-1}+ix_{2n}) \simeq (x_1, \dots, x_{2n}) = \\
                                          &&& ( r \cos(\theta) , r \sin(\theta) , \sqrt{1-r^2} \cos(\varphi_1), \sqrt{1-r^2} \sin(\varphi_1) \cos(\varphi_2) , \dots , \\
                                          &&& \sqrt{1-r^2} \sin(\varphi_1) \dots \cos(\varphi_{2n-3}), \sqrt{1-r^2} \sin(\varphi_1) \dots \sin(\varphi_{2n-3})  ). \end{array} \] 
This amounts to choose the modulus $r$ and the argument $\theta$ of $x_1+ix_2$, and then describe the remaining parameters using the standard spherical coordinates on $\mathbb{S}^{n-1}$, seen as a sphere of real dimension $2n-3$, including a radius factor $\sqrt{1-r^2}$. The advantage of these coordinates is that $C_{n}^{\varepsilon}(1, 0,\dots,0)$ simply corresponds to the set of points for which $r \geq \varepsilon$. 


The metric in these coordinates happens to be diagonal, given by:
\begin{itemize}
\item $g_{rr} = \braket{\vec{e_r} \vert \vec{e_r} } = 1 + \frac{r^2}{1-r^2}$
\item $g_{\theta \theta} = \braket{\vec{e_\theta} \vert \vec{e_\theta} } = r^2$
\item $g_{\varphi_i \varphi_i} = (1-r^2) [g_{\varphi_i \varphi_i}]$ with $[g]$ the metric corresponding to the spherical coordinates on $\mathbb{S}^{n-1}$.
\end{itemize}

It is now easy to compute the desired quantity:

\begin{align*}
A_n^{\varepsilon} &= \int_\varepsilon^1 \sqrt{1 + \frac{r^2}{1-r^2}} \mathrm{d}r \int_{[0,2\pi[\times [0,\pi[^{2n-4} \times [0,2\pi[ } r\mathrm{d}\theta \sqrt{1-r^2}^{2n-3} \sqrt{[g]} \mathrm{d}\varphi_1 \dots \mathrm{d}\varphi_{2n-3} \\
&= 2\pi A_{n-1} \int_\varepsilon^1 r(1-r^2)^{n-2} \mathrm{d}r \\
&= \frac{\pi A_{n-1}}{n-1} \int_\varepsilon^1 2(n-1)r(1-r^2)^{n-2} \mathrm{d}r \\
&= \frac{\pi A_{n-1}}{n-1} (1-\varepsilon^2)^{n-1}
\end{align*}
and, finally, 
\[ \frac{A_n^{\varepsilon}}{A_n} =  \frac{A_n^{\varepsilon}}{A_n^0} = (1-\varepsilon^2)^{n-1}.\]
\end{proof}

We are now ready to prove the theorem.

\begin{proof}[Proof of Theorem \ref{estimate eta}]
For this proof, we find some inspiration in \cite{moran1979closest}, but eventually obtain sharper bounds with simpler arguments. Another major reference concerning spherical caps is \cite{rankin1955closest}. We say that a set of vectors on a sphere are $\varepsilon$-separated if all scalar products between any pairs among them are not greater than $\varepsilon$ in modulus. Denote $\mathrm{d}\bar{\sigma}^n$ the normalized Lebesgue's measure on $\mathbb{S}^{n}$, that is $\mathrm{d}\bar{\sigma}^n = \frac{\mathrm{d}\sigma^n}{A_n}$, and consider the following events:
\begin{itemize}
\item $A : \forall k \in \llbracket 1 , d-1 \rrbracket, \lvert \braket{ \mathcal{E}_d \vert  \mathcal{E}_k } \rvert \leq \varepsilon$
\item $B :  (\ket{\mathcal{E}_k})_{1\leq k \leq d-1}$ are $\varepsilon-$separated
\end{itemize}
so as to write $\mathbb{P}(\eta_{n,d} \leq \varepsilon) = \mathbb{P}(A \mid B)  \mathbb{P}(B) = \frac{ \mathbb{P}(A \cap B)}{ \mathbb{P}(B)} \mathbb{P}(\eta_{n,d-1} \leq \varepsilon)$, with:

\begin{align*}
\frac{ \mathbb{P}(A \cap B)}{ \mathbb{P}(B)} &= \frac{ \displaystyle  \int_{(\mathbb{S}^{n})^{d-1}} \mathrm{d}\bar{\sigma}^n(x_1) \dots \mathrm{d}\bar{\sigma}^n(x_{d-1}) \bbone_{ \{ x_1, \dots, x_{d-1} \text{ are $\varepsilon$-separated} \} }  \left(1- \frac{\left\vert \bigcup_{k=1}^{d-1} C_{n}^{\varepsilon}(x_k)  \right\vert}{A_n}  \right) }{ \displaystyle \int_{(\mathbb{S}^{n})^{d-1}} \mathrm{d}\bar{\sigma}^n(x_1) \dots \mathrm{d}\bar{\sigma}^n(x_{d-1}) \bbone_{ \{ x_1, \dots, x_{d-1} \text{ are $\varepsilon$-separated} \} }  }  \\
&=  1-  \mathbb{E}\left( \frac{ \left\vert \bigcup_{k=1}^{d-1} C_{n}^{\varepsilon}(\ket{\mathcal{E}_k})  \right\vert}{A_n}  \Bigg\vert B  \right).
\end{align*}

We need to find bounds on the latter quantity. Obviously, $\mathbb{E}\left( \frac{ \left\vert \bigcup_{k=1}^{d-1} C_{n}^{\varepsilon}(\ket{\mathcal{E}_k})  \right\vert}{A_n}  \Bigg\vert B  \right) \leq (d-1) \frac{A_n^{\varepsilon}}{A_n}$, corresponding to the case when all the caps are disjoint. For the lower bound, define the sequence $u_d = \mathbb{E}\left( \frac{ \left\vert \bigcup_{k=1}^{d} C_{n}^{\varepsilon}(\ket{\mathcal{E}_k})  \right\vert}{A_n} \right)$, which clearly satisfies $u_d \leq \mathbb{E}\left( \frac{ \left\vert \bigcup_{k=1}^{d} C_{n}^{\varepsilon}(\ket{\mathcal{E}_k})  \right\vert}{A_n}  \Bigg\vert B  \right)$, because conditioning on the vectors being separated can only decrease the overlap between the different caps. First observe that $u_1 = \frac{A_n^{\varepsilon}}{A_n} \equiv \alpha$, and compute:

\begin{align*}
u_d &= u_{d-1} + \mathbb{E}\left(   \frac{ \left\vert C_{n}^{\varepsilon}(\ket{\mathcal{E}_d})  \setminus  \bigcup_{k=1}^{d-1} C_{n}^{\varepsilon}(\ket{\mathcal{E}_k})  \right\vert}{A_n}  \right) \\
&=  u_{d-1} +  \int_{(\mathbb{S}^{n})^{d}} \mathrm{d}\bar{\sigma}^n(x_1) \dots \mathrm{d}\bar{\sigma}^n(x_d) \int_{C_{n}^{\varepsilon}(x_d)} \bbone_{ \{ y \notin \bigcup_{k=1}^{d-1} C_{n}^{\varepsilon}(x_k) \} } \mathrm{d}\bar{\sigma}^n(y) \\
&=   u_{d-1} +  \int_{(\mathbb{S}^{n})^{d-1}} \mathrm{d}\bar{\sigma}^n(x_1) \dots \mathrm{d}\bar{\sigma}^n(x_{d-1}) \int_{\mathbb{S}^{n}} \frac{\left\vert C_{n}^{\varepsilon}(y) \right\vert}{A_n} \bbone_{ \{ y \notin \bigcup_{k=1}^{d-1} C_{n}^{\varepsilon}(x_k) \} } \mathrm{d}\bar{\sigma}^n(y) \\
&=  u_{d-1} + \frac{A_n^\varepsilon}{A_n}  \int_{(\mathbb{S}^{n})^{d-1}} \mathrm{d}\bar{\sigma}^n(x_1) \dots \mathrm{d}\bar{\sigma}^n(x_{d-1}) \left( 1- \frac{ \left\vert \bigcup_{k=1}^{d-1} C_{n}^{\varepsilon}(x_k)  \right\vert}{A_n} \right) \\
&= u_{d-1} + \frac{A_n^{\varepsilon}}{A_n} (1-u_{d-1}) \\
&= (1-\alpha)u_{d-1} + \alpha,
\end{align*} 
where the main trick was to invert the integrals on $x_d$ and on $y$. This result is actually quite intuitive: it states that when adding a new cap, only a fraction $1-u_{d-1}$ of it on average will be outside the previous caps and contribute to the new total area covered by the caps. Hence $u_d = 1- (1-\alpha)^d$, and the recurrence relation becomes: 
\[ \left(1 - (d-1) \frac{A_n^{\varepsilon}}{A_n} \right) \mathbb{P}(\eta_{n,d-1}\leq \varepsilon) \leq \mathbb{P}(\eta_{n,d} \leq \varepsilon) \leq \left(1-\frac{A_n^{\varepsilon}}{A_n} \right)^{d-1} \mathbb{P}(\eta_{n,d-1}\leq \varepsilon).\] 
Applying the lemma, we get by induction: 

\[ \prod_{k=1}^{d-1} (1-k (1-\varepsilon^2)^{n-1} ) \leq \mathbb{P}(\eta_{n,d} \leq \varepsilon) \leq (1- (1-\varepsilon^2)^{n-1})^{\frac{d(d-1)}{2} }. \]
Note that the left inequality is valid only as long as $d \leq \big( \frac{1}{1-\varepsilon^2} \big)^{n-1}$, but when $d$ is larger than this critical value, the right hand side becomes very small (of order $e^{-1/2(1-\varepsilon^2)^{n-1}}$), so we may take $0$ as a good lower bound in this case. The two bounds are in fact extremely close to each other, and get closer as $n$ or $d$ goes larger. To quantify this precisely, let's denote $f_{n,d}(\varepsilon) = (1- (1-\varepsilon^2)^{n-1})^{\frac{d(d-1)}{2}}$, $g_{n,d}(\varepsilon) = \prod_{k=1}^{d-1} (1-k (1-\varepsilon^2)^{n-1} )$, and let's show that $\lvert f_{n,d}(\varepsilon) - g_{n,d}(\varepsilon) \rvert \underset{n \text{ or } d \rightarrow \infty}{\longrightarrow} 0$. Two cases have to be considered.

\begin{itemize}
\item \textit{First case}: if $d \geq d_c \equiv \left(\frac{1}{1-\varepsilon^2}\right)^{\frac{3}{5} (n-1)}$, then $f_{n,d}(\varepsilon)$ is small so we can write:
\begin{align*}
 \lvert f_{n,d}(\varepsilon) - g_{n,d}(\varepsilon) \rvert &\leq f_{n,d}(\varepsilon) = e^{\frac{d(d-1)}{2} \ln(1- (1-\varepsilon^2)^{n-1})}  \\
&\leq e^{-\frac{(d-1)^2}{2} (1-\varepsilon^2)^{n-1}} \quad \quad \quad \text{(since $\ln(1-x) \leq -x$)} \\
 &\leq e^{ -\frac{1+ o(1)}{2 (1-\varepsilon^2)^{\frac{n-1}{5}}}} \quad \quad \quad \text{(using $d \geq d_c$)} \\
&\leq e^{-\frac{d^{1/3}}{2} (1+o(1))},
\end{align*}
where $1+o(1) = \left( \frac{d-1}{d} \right)^2 \underset{n \text{ or } d \rightarrow \infty}{\longrightarrow} 1$. \\

\item \textit{Second case}: if $d \leq d_c$, first note that $\forall k \in \llbracket 1,d \rrbracket, \forall x \in [ 0 ,\frac{1}{d^{5/3}}[$, 
\begin{align*} 1 \leq \frac{(1-x)^k}{1-kx} \leq \frac{1- kx + \frac{k(k-1)}{2}x^2}{1-kx} \leq 1+ \frac{k(k-1)}{2} \frac{x^2}{(1-x^{2/5})}. \end{align*}
Therefore, 
\begin{align*}
\left\lvert \ln(f_{n,d}(\varepsilon) - \ln(g_{n,d}(\varepsilon)) \right\rvert 
=& \left\lvert  \sum_{k=1}^{d-1} k \ln(1- (1-\varepsilon^2)^{n-1}) - \ln(1-k (1-\varepsilon^2)^{n-1}) \right\rvert \\
\leq&  \sum_{k=1}^{d-1} \left\lvert  \ln\left(1+ \frac{k(k-1)}{2} \frac{(1-\varepsilon^2)^{2(n-1)}}{1- (1-\varepsilon^2)^{\frac{2}{5}(n-1)}} \right)  \right\rvert   \quad \quad \text{(applying the inequality for $x = (1-\varepsilon^2)^{n-1}$)} \\
\leq& \frac{(1-\varepsilon^2)^{2(n-1)}}{1- (1-\varepsilon^2)^{\frac{2}{5}(n-1)}} \underbrace{\sum_{k=1}^{d-1}  \frac{k(k-1)}{2}}_{=\frac{d^3}{6} - \frac{d^2}{2} + \frac{d}{3} \leq \frac{d^3}{6} } \\
\leq& \frac{ (1-\varepsilon^2)^{\frac{n-1}{5}} }{ 6 (1-(1-\varepsilon^2)^{\frac{2}{5}(n-1)}) }  \quad \quad \quad \text{(using $d \leq d_c$)} \\
\leq&  \frac{d^{-\frac{1}{3}} }{ 6 (1-d^{-\frac{2}{3}}) }. \\ 
\end{align*}
Hence:
\begin{align*}
&\frac{g_{n,d}(\varepsilon)}{f_{n,d}(\varepsilon)} \in \Big[ \exp\left(-\frac{ (1-\varepsilon^2)^{\frac{n-1}{5}} }{ 6 (1-(1-\varepsilon^2)^{\frac{2}{5}(n-1)}) } \right), 1  \Big] \\
\Rightarrow \quad &\lvert f_{n,d}(\varepsilon) - g_{n,d}(\varepsilon) \rvert \leq \left(1 - \exp\left(- \frac{ (1-\varepsilon^2)^{\frac{n-1}{5}} }{ 6 (1-(1-\varepsilon^2)^{\frac{2}{5}(n-1)}) } \right) \right) f_{n,d}(\varepsilon) \\
& \hskip2.7cm \leq  \frac{ (1-\varepsilon^2)^{\frac{n-1}{5}} }{ 6 (1-(1-\varepsilon^2)^{\frac{2}{5}(n-1)}) } \quad \quad \quad \text{(since $1-e^{-x} \leq x$ and $f_{n,d}(\varepsilon) \leq 1$)} \\
& \hskip2.7cm \leq  \frac{d^{-\frac{1}{3}} }{ 6 (1-d^{-\frac{2}{3}}) }.
\end{align*}
\end{itemize}

We have thus shown that the difference between the two bounds $f_{n,d}(\varepsilon)$ and $g_{n,d}(\varepsilon)$ can be controlled by a quantity that can be expressed solely in terms of either $n$ or $d$ but that anyway vanishes when either $n$ or $d$ tend to infinity. If we call $\xi$ this vanishing term, it is straightforward to see that:
 \[\min \{ d \in \mathbb{N} \mid 1- f_{n,d}(\varepsilon) + \xi \geq s \}  \leq d_{max}^{\varepsilon, s}(n) \leq \min \{ d \in \mathbb{N} \mid 1- f_{n,d}(\varepsilon) \geq s \}, \]
and after some work, this implies:
\[\left\lfloor \sqrt{-2\ln(1-s+\xi)} \sqrt{\frac{1}{(1-\varepsilon^2)^{n-1}} -1} \right\rfloor \leq  d_{max}^{\varepsilon, s}(n) \leq \left\lceil \frac{\sqrt{-2\ln(1-s)} }{(1-\varepsilon^2)^{\frac{n-1}{2}}} \right\rceil, \]
hence $\displaystyle d_{max}^{\varepsilon, s}(n) \underset{n \rightarrow \infty}{\sim} \sqrt{-2 \ln(1-s)} \left( \frac{1}{1-\varepsilon^2} \right)^{\frac{n-1}{2}}$, which is the first part of the theorem.


The intuition concerning the second statement comes from the following observation. We know that $\mathbb{P}(\eta_{n,d} \leq \varepsilon) \simeq f_{n,d}(\varepsilon)$, and this function happens to be almost constant equal to $0$ in the vicinity of $\varepsilon=0$, almost $1$ in the vicinity of $\varepsilon=1$, and to have a very sharp step between the two; this step sharpens as $n$ or $d$ grows larger. This explains why the mass of probability is highly peaked around a critical value $\varepsilon_c$, so that $\displaystyle \eta_{n,d} \simeq \mathbb{E}(\eta_{n,d})$ converges to a deterministic variable when $n$ or $d$ $\rightarrow \infty$. This is certainly due to the averaging effect of considering the maximum of a set of $\frac{d(d+1)}{2}$ scalar products. The critical value $\varepsilon_c$ satisfies:
\[ (1- (1-\varepsilon_c^2)^{n-1})^{\frac{d(d-1)}{2}} = \frac{1}{2} \Leftrightarrow \varepsilon_c = \sqrt{ 1- (1- 2^{-2/d(d-1)})^{1/n-1} } \simeq \sqrt{1-d^{-2/n}}. \]

Now, the precise proof of the convergence of $\eta_{n,d}$ in probability goes as follows. Let $\delta>0$. We have to show that $\mathbb{P}\left( \left\lvert \eta_{n,d} - \sqrt{1-d^{-\frac{2}{n}}} \right\rvert \leq \delta \right) \underset{n \text{ or } d \rightarrow \infty}{\longrightarrow} 1$. It is equivalent but easier to show that $\mathbb{P}\left(\sqrt{1-d^{-\frac{2}{n}} - \delta} \leq \eta_{n,d} \leq \sqrt{1-d^{-\frac{2}{n}} +\delta} \right) \underset{n \text{ or } d \rightarrow \infty}{\longrightarrow} 1$. Taking $f_{n,d}$ as an approximation for the distribution function of $\eta_{n,d}$, we can write: 
\[ \mathbb{P}\left(\eta_{n,d} \leq \sqrt{1-d^{-\frac{2}{n}} +\delta} \right) = \left(1- \max\left(0,d^{-2/n} - \delta \right)^{n-1} \right)^{\frac{d(d-1)}{2}} + o(1), \]
where $o(1)$ stands for a quantity that goes to zero when either $n$ or $d$ goes to infinity (bounded by $\xi$), and where the max appears because if $d^{-2/n} \leq \delta$, $\mathbb{P}\left(\eta_{n,d} \leq \sqrt{1-d^{-\frac{2}{n}} +\delta} \right)$ is simply equal to 1. Clearly, $\mathbb{P}\left(\eta_{n,d} \leq \sqrt{1-d^{-\frac{2}{n}} +\delta} \right) \underset{n \text{ or } d \rightarrow \infty}{\longrightarrow} 1$, and similarly, one shows that $\mathbb{P}\left(\eta_{n,d} \leq \sqrt{1-d^{-\frac{2}{n}} - \delta} \right) \underset{n \text{ or } d \rightarrow \infty}{\longrightarrow} 0$, which completes the proof.
\end{proof}

During the reviewing process, we discovered that the formula $\sqrt{1-d^{-\frac{2}{n}}}$ had already been obtained in \cite{zhang2017spherical}. However, this work only deals with the maximum of the $d$ scalar products between say the north pole and a set of $d$ independent random vectors. This situation is easier to treat, in particular because the $d$ scalar products are then independent random variables, which is certainly not the case for our $\frac{d(d+1)}{2}$ scalar products.

\subsection{Comparison with simulation and consequences} \label{consequences}

\begin{figure}[h]
\hfill
\subfigure[$\varepsilon=0.1$ and $s=0.9$]{\includegraphics[scale=0.45]{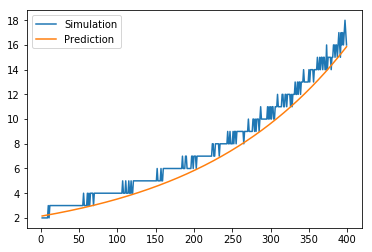}}
\hfill
\subfigure[$\varepsilon=0.4$ and $s=0.9$]{\includegraphics[scale=0.45]{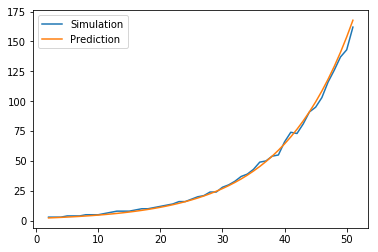}}
\hfill
\caption{Simulation \textit{vs} prediction for $d_{max}^{\varepsilon, s}(n)$}
\label{fig1}
\end{figure}

\begin{figure}[h]
\hfill
\subfigure[$n=10$]{\includegraphics[scale=0.45]{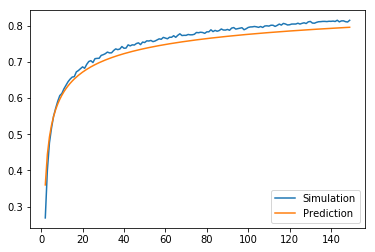}}
\hfill
\subfigure[$n=50$]{\includegraphics[scale=0.45]{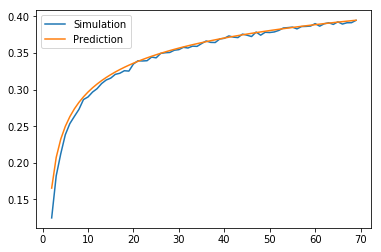}}
\hfill
\caption{Simulation \textit{vs} prediction for $d \mapsto \mathbb{E}(\eta_{n,d})$ at fixed $n$}
\label{fig2}
\end{figure}

The above expressions actually give incredibly good estimations for $d_{max}^{\varepsilon, s}(n)$ and $\eta_{n,d}$, as shown in Figures \ref{fig1} and \ref{fig2}. 

This theorem has a strong physical consequence. Indeed, $\mathcal{E}$ induces proper decoherence on $\mathcal{S}$ as long as $\eta_{n,d} \ll 1$, that is when $d^{-2/n}$ is very close to $1$, \textit{i.e.} when $d \ll e^{n/2}$. Going back to physically meaningful quantities, we write as previously $n = p^{N_\mathcal{E}}$ and $d = p^{N_\mathcal{S}}$ where $N_\mathcal{E}$ and $N_\mathcal{S}$ stand for the number of particles composing $\mathcal{E}$ and $\mathcal{S}$. The condition becomes: $2 \ln(p)N_\mathcal{S}  \ll p^{N_\mathcal{E}}$ or simply:
 \[ \boxed{ \frac{\ln(N_\mathcal{S})}{\ln(p)}  \ll N_\mathcal{E}. }\]
A more precise condition can be obtained using $d_{max}$, because $\mathcal{E}$ induces proper decoherence on $\mathcal{S}$ as long as $d \leq d_{max}^{\varepsilon, s}(n)$ for an arbitrary choice of $\varepsilon$ close to $0$ and $s$ close to $1$. This rewrites: $2\ln(p)N_\mathcal{S} \leq \ln(\sqrt{-2 \ln(1-s)}) + \ln \left( \frac{1}{1-\varepsilon^2} \right) p^{N_\mathcal{E}} \simeq \varepsilon^2 p^{N_\mathcal{E}}$ or simply: $\ln(N_\mathcal{S}) \leq 2 \ln(\varepsilon) + \ln(p) N_\mathcal{E}$. Thus, for instance, a gas composed of thousands of particles will lose most of its coherence if it interacts with only a few external particles. It is rather surprising that so many points can be placed randomly on a $n$-sphere before having the maximum of the scalar products becoming non-negligible. \textit{It is this property that makes decoherence an extremely efficient high-dimensional geometrical phenomenon}. 

\subsection{Discussing the hypotheses} \label{hypotheses}

On the one hand, this result could be seen as a worst case scenario for decoherence, since realistic Hamiltonians are far from random and actually discriminate even better the different possible histories. This is especially true if $\mathcal{E}$ is a measurement apparatus for example, whose Hamiltonian is by construction such that the $(\ket{\mathcal{E}_i(t)})_{1\leq i \leq d}$ evolve quickly and deterministically towards orthogonal points of the sphere.

On the other hand, pursuing such a high level of generality led us to abstract and unphysical assumptions. First, realistic dynamics are not isotropic on the $n$-sphere (some transitions are more probable than others). Then, the assumption that each $\ket{\mathcal{E}_i(t)}$ can explore indistinctly all the states of $\mathcal{H}_{\mathcal{E}}$ is very criticizable. As explained in \cite{poulin2011quantum}:

\begin{quotation}
‘\dots the set of quantum states that can be reached from a product state with a polynomial-time evolution of an arbitrary time-dependent quantum Hamiltonian is an exponentially small fraction of the Hilbert space. This means that the vast majority of quantum states in a many-body system are unphysical, as they cannot be reached in any reasonable time. As a consequence, all physical states live on a tiny submanifold.’
\end{quotation}

It would then be more accurate in our model to replace $\mathbb{S}^{n}$ by this submanifold. But how does it look like geometrically and what is its dimension? If it were a subsphere of $\mathbb{S}^{n}$ of exponentially smaller dimension, then $n$ should be replaced everywhere by something like $\ln(n)$ in what precedes, so the condition would rather be $N_\mathcal{S} \ll N_\mathcal{E}$ which is a completely different conclusion. Some clues to better grasp the submanifold are found in \cite[\S3.4]{orus2014practical}: 

\begin{quotation}
‘\dots one can prove that low-energy eigenstates of gapped Hamiltonians with local interactions obey the so-called area-law for the entanglement entropy. This means that the entanglement entropy of a region of space tends to scale, for large enough regions, as the size of the boundary of the region and not as the volume. (\dots) In other words, low-energy states of realistic Hamiltonians are not just “any” state in the Hilbert space: they are heavily constrained by locality so that they must obey the entanglement area-law.’
\end{quotation}
More work is needed in order to draw precise conclusions taking this physical remarks into account.

\section{Second model: interacting particles} \label{second model}
\subsection{The environment feels the system} \label{feels}

At present, let's better specify the nature of the environment. Suppose that the energy of interaction dominates the evolution of the whole system $\mathcal{S} + \mathcal{E}$ and can be expressed in terms of the positions $x_1, \dots , x_N$ of the $N$ particles composing the environment, together with the state of $\mathcal{S}$ (this is the typical regime for macroscopic systems which decohere in the position basis \cite[\S III.E.2.]{schlosshauer2005decoherence}). If the latter is $\ket{i}$, denote $H(i , x_1 \dots x_N)$ this energy. The initial state $\ket{\Psi} = \left( \sum_{i=1}^d c_i \ket{i} \right) \otimes \int f( x_1 \dots  x_N) \ket{ x_1 \dots  x_N} \mathrm{d}x_1 \dots  \mathrm{d}x_N$ evolves into: 

 \[ \sum_{i=1}^d c_i \ket{i} \otimes \underbrace{\int  f( x_1 \dots  x_N) e^{\frac{i}{\hbar} H(i , x_1 \dots  x_N) t} \ket{ x_1 \dots x_N} }_{= \ket{\mathcal{E}_i(t)}} \mathrm{d}x_1 \dots  \mathrm{d}x_N. \]
Therefore:
\[ \braket{ \mathcal{E}_i(t) \vert \mathcal{E}_j(t)}  = \int \lvert f( x_1 \dots  x_N) \rvert^2 e^{\frac{i}{\hbar} \Delta(i,j, x_1 \dots x_N) t}  \mathrm{d}x_1 \dots  \mathrm{d}x_N, \]
where $\Delta(i,j, x_1 \dots x_N) = H(j , x_1 \dots  x_N) - H(i , x_1 \dots  x_N)$ is a spectral gap between eigenvalues of the Hamiltonian, measuring how much the environment feels the transition of $\mathcal{S}$ from $\ket{i}$ to $\ket{j}$ in a given configuration of the environment. In a time interval $[-T,T]$, the mean value $\frac{1}{2T} \int_{-T}^T \braket{ \mathcal{E}_i(t) \vert \mathcal{E}_j(t)} \mathrm{d}t$ yields $ \int \lvert f( x_1 \dots  x_N) \rvert^2 \mathrm{sinc}(\frac{\Delta(i,j, x_1 \dots x_N)}{\hbar} T)$ which is close to zero for all $i$ and $j$ as soon as $T > \frac{\pi \hbar}{\displaystyle \min_{i, j, x_1 \dots x_N} \Delta(i,j, x_1 \dots x_N)}$, which is likely to be small if $\mathcal{E}$ is a macroscopic system, for the energies involved will be much greater than $\hbar$. Similarly, the empirical variance is: 
\[ \mathbb{V} = \frac{1}{2T} \int_{-T}^T \lvert \braket{ \mathcal{E}_i(t) \vert \mathcal{E}_j(t)} \rvert^2 \mathrm{d}t \sim \int \lvert f( x_1 \dots  x_N) \rvert^4 \mathrm{d}x_1 \dots  \mathrm{d}x_N, \]
 plus terms that go to zero after a short time. Note that the variables $x_1 \dots  x_N$ could be discretized to take $p$ possible values, in which case $n = \dim(\mathcal{H}_\mathcal{E}) = p^N$, and the integral becomes a finite sum. For a delocalized initial state with constant $f$, this sum is equal to $p^{-N}$, and we recover the previous estimate \eqref{small_eta} if $d=2$: $\eta \sim p^{-N/2}$. This model teaches us that \textit{the more the environment feels the difference between the possible histories, the more they decohere}. 

\subsection{Entanglement entropy as a measure of decoherence} \label{entropy}

What precedes suggests the following intuition: the smaller $\eta$ is, the more information the environment has stored about the system because the more distinguishable (\textit{i.e.} orthogonal) the $(\ket{\mathcal{E}_i(t)})_{1\leq i \leq d}$ are; on the other hand, the smaller $\eta$ is, the fewer quantum interferences occur. It motivates the search for a general relationship between entanglement entropy (defined as the von Neumann entropy of the reduced density matrix of either $\mathcal{S}$ or $\mathcal{E}$, \textit{i.e.} how much $\mathcal{E}$ knows about $\mathcal{S}$ or vice-versa) and the level of classicality of a system. Such results have already been derived for specific environments \cite[(3.76)]{joos1996decoherence} \cite{le2008entanglement} \cite{merkli2018production} but not, to our knowledge, in the general case. The following formula is proved in the annex \ref{annex} when $S$ stands for the linear entropy (or purity defect) $1-\tr(\rho^2)$, and some justifications are given when $S$ denotes the entanglement entropy: 

\begin{equation} 
\forall F \subset \mathcal{H}_{\mathcal{S}}, \quad \lvert \tr(\rho_{\mathcal{S}}(t) \Pi_F) -  \tr(\rho_{\mathcal{S}}^{(d)} \Pi_F) \rvert \leq \dim(F) \; \sqrt{ 1 -  \inf_{\ket{\Psi_\mathcal{S}(0)}} \frac{ S(\rho_{\mathcal{S}}(t)) } {S(\rho_{\mathcal{S}}^{(d)})} }. \label{nice_formula} 
\end{equation}

\section{Alternative definitions for $\eta$} \label{alternative}
Lemma \ref{lemma} allows for another way to quantify decoherence, which could be to ask, for each possible history $i$, what is the fraction $F^\varepsilon_{n,d}$ of the other possible histories with which it interferes significantly, that is how many indices $j$ are there such that $\lvert \braket{\mathcal{E}_i \vert \mathcal{E}_j} \rvert \geq \varepsilon$? As remarked in \cite{wyner1967random}, this quantity is simply given by:
\[ F^\varepsilon_{n,d} = \frac{1}{d-1} \sum_{\substack{j=1 \\ j \neq i}}^d h_{ij}  \quad \text{with } h_{ij} = 
 \left\{ \begin{array}{ll} 1 \text{ if } \ket{ \mathcal{E}_j} \in C^\varepsilon_n(\ket{\mathcal{E}_i})  \\  0 \text{ otherwise } \end{array} \right. \] 
By the law of large numbers, we immediately deduce that $F^\varepsilon_{n,d}  \underset{d \rightarrow \infty}{\overset{\text{a.s.}}{ \longrightarrow}} \mathbb{P}(h_{ij} = 1) = (1- \varepsilon^2)^{n-1}$, and we recover once again in this expression that the typical level of decoherence is $\varepsilon \sim \frac{1}{\sqrt{n}}$.

More interesting, perhaps, could be to quantify decoherence using the ‘expectation’ of the scalar products $\lvert \braket{\mathcal{E}_i \vert \mathcal{E}_j} \rvert^2$ for $i\neq j$, weighted by their quantum probabilities $\frac{\lvert c_i \rvert^2\lvert c_j \rvert^2}{\sum_{i\neq j} \lvert c_i \rvert^2\lvert c_j \rvert^2} = \frac{\lvert c_i \rvert^2\lvert c_j \rvert^2}{1 - \sum_i \lvert c_i \rvert^4}$. One could then define: 
\[ \tilde{\eta}^2 = \frac{1}{1 - \sum_i \lvert c_i \rvert^4} \sum_{i \neq j} \lvert c_i \rvert^2\lvert c_j \rvert^2 \lvert \braket{\mathcal{E}_i \vert \mathcal{E}_j} \rvert^2 =  \frac{1}{1 - \sum_i \lvert c_i \rvert^4} \left( \tr(\rho^2_{\mathcal{S}}) - \sum_i \lvert c_i \rvert^4 \right), \]
based on a computation made in the annex \ref{linear entropy}. As always, decoherence depends on the basis considered. This is clear in the above definition, including a basis-independent (‘covariant’) part $\tr(\rho^2_{\mathcal{S}})$, but also the term $\sum_i \lvert c_i \rvert^4$ depending the diagonal elements of $\rho_{\mathcal{S}}$, which are basis-dependent.

This definition has two advantages. First, it can naturally be extended to the infinite dimensional case, unlike $\displaystyle \max_{i \neq j} \; \lvert \braket{\mathcal{E}_i(t) \vert \mathcal{E}_j(t)} \rvert$ (if the scalar products vary continuously, their supremum is necessarily 1). A proposal for quantifying decoherence in infinite dimension is therefore:
\[ \tilde{\eta}^2 =  \frac{1}{1 - \int \lvert c_x \rvert^4 \mathrm{d}x} \left( \tr(\rho^2_{\mathcal{S}}) - \int \lvert c_x \rvert^4 \mathrm{d}x \right). \]
Second, it allows to give a physical meaning to the purity of a state: $\tr(\rho^2_{\mathcal{S}})$ \textit{can be seen as the typical level of decoherence in most bases}. To see this, start from $\rho_{\mathcal{S}}$'s eigenbasis, denoted $\mathcal{B}_e = (\ket{e_i})_{1\leq i \leq d}$, in which $\rho_{\mathcal{S}} =  \begin{pmatrix} \lambda_1 & &  \\  & \ddots & \\  & &  \lambda_d  \end{pmatrix}$ and $\tilde{\eta}_{\mathcal{B}_e}^2 = 0$. Now, pick randomly another basis; more precisely, choose a unitary $U$ according to the Haar measure on $\mathcal{U}(\mathcal{H}_\mathcal{S})$, yielding the new basis $\mathcal{B}' = (U\ket{e_i})_{1\leq i \leq d}$. Expressed in $\mathcal{B}'$, the i$^{th}$ diagonal coefficient of $\rho_{\mathcal{S}}$ is:
$\lvert c'_i \rvert^2 = \braket{ U e_i \vert \rho_\mathcal{S} \vert U e_i } = \braket{Ue_i \vert \sum_k \lambda_k \ket{e_k}\bra{e_k} \vert Ue_i} = \sum_k \lambda_k \lvert \braket{e_i \vert U e_k} \rvert^2$. Since $U$ follows a uniform law in $\mathcal{U}(\mathcal{H}_\mathcal{S})$, so does $\ket{U e_k}$ on $\mathbb{S}^{d}$. Therefore, according to Proposition \ref{variance S}, each $\lvert \braket{e_i \vert U e_k} \rvert^2$ is expected to be close to $\frac{1}{d}$, and this estimate will be even more precise for their weighted mean value, hence $\lvert c'_i \rvert^2 \simeq \frac{1}{d}$ and $\sum_i \lvert c'_i \rvert^4 \simeq \frac{1}{d}$. Consequently, in a typical random basis, $\tilde{\eta}_{\mathcal{B'}}^2 \simeq \frac{1}{1 - \sum_i \frac{1}{d}} \left( \tr(\rho^2_{\mathcal{S}}) - \frac{1}{d} \right)$. For example, when $\rho_{\mathcal{S}}$ is a maximally entangled state, $\tr(\rho^2_{\mathcal{S}}) = \frac{1}{d}$ and we deduce that $\tilde{\eta}^2 = 0$ (perfect decoherence) in most bases. In this case it is actually 0 in \textit{all} bases, because $\rho_{\mathcal{S}}$ is a scalar matrix, diagonal in any basis. For a generic state however, if $d$ is sufficiently high, one can expect $\tr(\rho^2_{\mathcal{S}}) \gg \frac{1}{d}$. Therefore, in most bases, $\boxed{\tilde{\eta}^2 \simeq \tr(\rho^2_{\mathcal{S}})}$.


\section*{Conclusion}

We introduced, in a mathematically rigorous way, general quantities that can be relevant for any study on decoherence, in particular the parameter $\eta(t)$ that quantifies the level of decoherence at a given instant. Two simple models were then presented, designed to feel more intuitively the general process of decoherence. Most importantly, our study revealed the mathematical reason why the latter is so fast and universal, namely because surprisingly many points can be placed randomly on a $n$-sphere before having the maximum of the scalar products becoming non-negligible. We also learned that decoherence is neither perfect nor everlasting, since $\eta$ is not expected to be exactly 0 and will eventually become large again (according to Borel-Cantelli's lemma for the first model, and finding particular times such that all the exponentials are almost real in the second) pretty much like the ink drop in the glass of water will re-form again due to Poincaré's recurrence theorem, even though the recurrence time can easily exceed the lifetime of the universe for realistic systems \cite{zurek1982environment}. Finally, decoherence can be estimated by entanglement entropy because $\eta$ is linked to what the environment knows about the system.

Further works could include the search for a description of the submanifold of reachable states mentioned in \S\ref{hypotheses}; a generalization to the cases where the initial environment is not in a pure state; the study of the infinite dimensional case.

\begin{appendices}
\section{Annex: decoherence estimated by the entanglement entropy with the environment} \label{annex}
We establish here the formula \eqref{nice_formula}: we first derive the inequality \eqref{classicality}, and then look for a relation between $\eta$ and the linear entropy or the entanglement entropy. Inserting the second into the first directly yields \eqref{nice_formula}.

\subsection{Relation between $\eta$ and the level of classicality}
Let's keep the notations of \S\ref{basics}, where we defined $\rho_{\mathcal{S}}^{(q)}(t) = \sum_{i \neq j} c_i \bar{c_j}  \braket{\mathcal{E}_j(t) \vert \mathcal{E}_i(t)} \ket{i}\bra{j}$. We have $\triplenorm{ \rho_{\mathcal{S}}^{(q)}(t)} \leq \eta(t) $ because for all vectors $\ket{\Psi} = \sum_k \alpha_k \ket{k} \in \mathcal{H}_{\mathcal{S}}$ of norm 1, 
\begin{align*}
 & \rho_{\mathcal{S}}^{(q)}(t) \ket{\Psi} =  \sum_{1 \leq i \neq j \leq d} c_i \bar{c_j} \braket{\mathcal{E}_j(t) \vert \mathcal{E}_i(t)} \alpha_j \ket{i} \\
\Rightarrow &  \quad \lVert  \rho_{\mathcal{S}}^{(q)}(t) \ket{\Psi} \rVert^2 = \sum_{i=1}^d \lvert c_i \rvert^2 \lvert \sum_{\substack{j=1 \\ j \neq i}}^d \bar{c_j}  \braket{\mathcal{E}_j(t) \vert \mathcal{E}_i(t)} \alpha_j \rvert^2 \leq \eta(t)^2 \sum_{i=1}^d \lvert c_i \rvert^2  \sum_{j=1}^d \lvert c_j \rvert^2 \leq \eta(t)^2. 
\end{align*}
Now, if $F$ is a subspace of $\mathcal{H}_{\mathcal{S}}$ (\textit{i.e.} a probabilistic event), let $(\varphi_k)_k$ be an orthonormal basis of $F$. We have:
\begin{align*}
& \tr(\rho_{\mathcal{S}}(t) \Pi_F) -  \tr(\rho_{\mathcal{S}}^{(d)} \Pi_F) = \tr(\rho_{\mathcal{S}}^{(q)}(t) \Pi_F)  = \sum_{k=1}^{\dim(F)} \braket{\varphi_k \vert \rho_{\mathcal{S}}^{(q)}(t) \varphi_k} \\
\Rightarrow \quad & \lvert \tr(\rho_{\mathcal{S}}(t) \Pi_F) -  \tr(\rho_{\mathcal{S}}^{(d)} \Pi_F) \rvert \leq \sum_{k=1}^{\dim(F)} \triplenorm{\rho_{\mathcal{S}}^{(q)}(t)} \leq \dim(F) \eta(t).
\end{align*} 
In a nutshell: $\mathbb{P}_{\text{quantum}} = \mathbb{P}_{\text{classical}} + \mathcal{O}(\eta)$.

\subsection{Relation between $\eta$ and the linear entropy} \label{linear entropy}
We define the linear entropy (or purity defect) of a state $\rho$ to be $S_{\text{lin}}(\rho) = 1 - \tr(\rho^2)$. Since $\mathcal{S}$ is initially in a pure state, the quantity $\frac{S_{\text{lin}}(\rho_{\mathcal{S}}(t))}{ S_{\text{lin}}(\rho_{\mathcal{S}}^{(d)})}$ goes from 0 at $t=0$ to almost 1 when $t \rightarrow +\infty$. It measures the ratio of purity that has already been lost compared to its final ideal value. Recall that $\rho_{\mathcal{S}}(t) =  \sum_{i=1}^d \lvert c_i \rvert^2 \ket{i}\bra{i} +  \sum_{1 \leq i \neq j \leq d} c_i \bar{c_j}  \braket{\mathcal{E}_j(t) \vert \mathcal{E}_i(t)} \ket{i}\bra{j}$, so that:

\begin{align*}
\frac{S_{\text{lin}}(\rho_{\mathcal{S}}(t))} {S_{\text{lin}}(\rho_{\mathcal{S}}^{(d)})} &= \frac{1 - \sum_i \lvert c_i \rvert^4 -  \sum_{i \neq j} \lvert c_i \rvert^2 \lvert c_j \rvert^2  \lvert \braket{\mathcal{E}_i(t) \vert \mathcal{E}_j(t)} \rvert^2 }{1 - \sum_i \lvert c_i \rvert^4} \\
& \geq 1 - \eta^2(t) \frac{ \sum_{i \neq j} \lvert c_i \rvert^2 \lvert c_j \rvert^2 }{1 - \sum_i \lvert c_i \rvert^4} \\
& \geq 1 - \eta^2(t),
\end{align*}
since the last fraction always equals 1 because $1 = (\sum_i \lvert c_i \rvert^2)(\sum_i \lvert c_i \rvert^2) = \sum_i \lvert c_i \rvert^4 + \sum_{i \neq j} \lvert c_i \rvert^2 \lvert c_j \rvert^2$. Note that, for any given time $t$, this inequality is actually an equality for the initial state $\ket{\Psi_{\mathcal{S}}(0)} = c_{i_0} \ket{i_0} + c_{j_0} \ket{j_0}$ where $i_0$ and $j_0$ denote two indices such that $\eta(t) = \lvert \braket{\mathcal{E}_{i_0}(t) \vert \mathcal{E}_{j_0}(t)} \rvert$. Thus:
\[ \eta(t) = \sqrt{ 1 -  \inf_{\ket{\Psi_\mathcal{S}(0)}} \frac{ S_{\text{lin}}(\rho_{\mathcal{S}}(t)) } {S_{\text{lin}}(\rho_{\mathcal{S}}^{(d)})} }. \]

\subsection{Relation between $\eta$ and the entanglement entropy}
The entanglement entropy is always much harder to manipulate. We were not able to prove in the general case a similar result when the linear entropy $S_{\text{lin}}$ is replaced by the entanglement entropy $S$, but numerical simulations tend to indicate that the same formula is still (almost) true and that there exists a deep link between the quantity $1 - \eta^2(t)$ and the ratio $\frac{S(\rho_{\mathcal{S}}(t))}{S(\rho_{\mathcal{S}}^{(d)})}$. Here are some considerations to get convinced.

In dimension $d=2$, if one denotes $f(t) = \braket{\mathcal{E}_2(t) \vert \mathcal{E}_1(t)}$, one can write $\rho_{\mathcal{S}}(t) =  \begin{pmatrix} \lvert c_1 \rvert^2 & c_1 \bar{c_2} f(t) \\ \bar{c_1} c_2 \bar{f}(t) & \lvert c_1 \rvert^2 \end{pmatrix}$, whose eigenvalues are $\text{ev}_{\pm} = \frac{1}{2}(1 \pm \sqrt{( \lvert c_1 \rvert^2 -  \lvert c_2 \rvert^2)^2 + 4 f^2(t)  \lvert c_1 \rvert^2  \lvert c_2 \rvert^2 })$. At large times, $f \ll 1$ and after some calculations we get at leading order:
\[  \frac{S(\rho_{\mathcal{S}}(t))}{S(\rho_{\mathcal{S}}^{(d)})} = \frac{\text{ev}_{+}\ln(\text{ev}_{+}) + \text{ev}_{-}\ln(\text{ev}_{-}) }{ \lvert c_1 \rvert^2 \ln( \lvert c_1 \rvert^2) + \lvert c_2 \rvert^2 \ln( \lvert c_2 \rvert^2) }  \simeq 1 + \frac{\frac{\lvert c_1 \rvert^2  \lvert c_2 \rvert^2}{ \lvert c_1 \rvert^2 -  \lvert c_2 \rvert^2} (\ln( \lvert c_1 \rvert^2) - \ln( \lvert c_2 \rvert^2))  }{ \lvert c_1 \rvert^2 \ln( \lvert c_1 \rvert^2) + \lvert c_2 \rvert^2 \ln( \lvert c_2 \rvert^2)} \eta^2(t).  \]
The ratio preceding $\eta^2$ is a one-parameter real function in $\lvert c_1 \rvert^2$ (since $\lvert c_2 \rvert^2 = 1 - \lvert c_1 \rvert^2$) defined on $[0,1]$; it turns out that is takes only values in $[-1 , -0.7]$ and tends to $-1$ only when $\lvert c_1 \rvert^2$ tends to $0$ or $1$. Therefore, in dimension 2, we still have (at least at leading order):
\[ \eta(t) = \sqrt{ 1 -  \inf_{\ket{\Psi_\mathcal{S}(0)}} \frac{ S(\rho_{\mathcal{S}}(t)) } {S(\rho_{\mathcal{S}}^{(d)})} }. \]

In higher dimension, if we suppose that one of the $\braket{\mathcal{E}_j(t) \vert \mathcal{E}_i(t)}$ decreases much slower than the others (assume without loss of generality that it is $\braket{\mathcal{E}_2(t) \vert \mathcal{E}_1(t)}$, still denoted $f(t)$), then after some time $\rho_{\mathcal{S}}(t)$ is not very different from: 

\[ \begin{pmatrix} \lvert c_1 \rvert^2 & c_1 \bar{c_2} f(t) & & & \\ \bar{c_1} c_2 \bar{f}(t) & \lvert c_1 \rvert^2 & & & \\ & & \lvert c_3 \rvert^2 & & \\& & & \ddots & \\ & & & & \lvert c_d \rvert^2 \end{pmatrix}. \]
Using the previous inequality in dimension 2:
\[ \frac{ S(\rho_{\mathcal{S}}(t)) } {S(\rho_{\mathcal{S}}^{(d)})} \geq \frac{ (1 - \eta^2(t))(\lvert c_1 \rvert^2 + \lvert c_2 \rvert^2 ) + \lvert c_3 \rvert^2 + \dots + \lvert c_d \rvert^2  }{\lvert c_1 \rvert^2 + \dots + \lvert c_d \rvert^2 } \geq 1 - \eta^2(t), \] 
and, once again, this bound is attained for an appropriate choice of the $(c_i)_{1\leq i \leq d}$.
\end{appendices}

\bibliographystyle{siam}
\bibliography{Biblio_decoherence}

\begin{thebibliography}{10}

\bibitem{di2021stable}
{\sc A.~Di~Biagio and C.~Rovelli}, {\em Stable facts, relative facts},
  Foundations of Physics, 51 (2021), pp.~1--13.
\newblock doi : 10.1007/s10701-021-00429-w.

\bibitem{fonda1978decay}
{\sc L.~Fonda, G.~Ghirardi, and A.~Rimini}, {\em Decay theory of unstable
  quantum systems}, Reports on Progress in Physics, 41 (1978), p.~587.
\newblock doi : 10.1088/0034-4885/41/4/003.

\bibitem{joos1996decoherence}
{\sc E.~Joos}, {\em Decoherence through interaction with the environment}, in
  Decoherence and the appearance of a classical world in quantum theory,
  Springer, 1996, pp.~35--136.
\newblock doi: 10.1007/978-3-662-03263-3.

\bibitem{le2008entanglement}
{\sc K.~Le~Hur}, {\em Entanglement entropy, decoherence, and quantum phase
  transitions of a dissipative two-level system}, Annals of Physics, 323
  (2008), pp.~2208--2240.
\newblock doi : 10.1016/j.aop.2007.12.003.

\bibitem{merkli2018production}
{\sc M.~Merkli, G.~Berman, R.~Sayre, X.~Wang, and A.~I. Nesterov}, {\em
  Production of entanglement entropy by decoherence}, Open Systems \&
  Information Dynamics, 25 (2018), p.~1850001.
\newblock doi : 10.1142/S1230161218500014.

\bibitem{moran1979closest}
{\sc P.~Moran}, {\em The closest pair of n random points on the surface of a
  sphere}, Biometrika, 66 (1979), pp.~158--162.
\newblock doi : 10.1093/BIOMET/66.1.158.

\bibitem{orus2014practical}
{\sc R.~Or{\'u}s}, {\em A practical introduction to tensor networks: Matrix
  product states and projected entangled pair states}, Annals of physics, 349
  (2014), pp.~117--158.
\newblock doi : 10.1016/j.aop.2014.06.013.

\bibitem{poulin2011quantum}
{\sc D.~Poulin, A.~Qarry, R.~Somma, and F.~Verstraete}, {\em Quantum simulation
  of time-dependent {H}amiltonians and the convenient illusion of {H}ilbert
  space}, Physical review letters, 106 (2011), p.~170501.
\newblock doi : 10.1103/PhysRevLett.106.170501.

\bibitem{rankin1955closest}
{\sc R.~A. Rankin}, {\em The closest packing of spherical caps in n
  dimensions}, Glasgow Mathematical Journal, 2 (1955), pp.~139--144.

\bibitem{saloff1994precise}
{\sc L.~Saloff-Coste}, {\em Precise estimates on the rate at which certain
  diffusions tend to equilibrium}, Mathematische Zeitschrift, 217 (1994),
  pp.~641--677.
\newblock doi : 10.1007/BF02571965.

\bibitem{schlosshauer2005decoherence}
{\sc M.~Schlosshauer}, {\em Decoherence, the measurement problem, and
  interpretations of quantum mechanics}, Reviews of Modern physics, 76 (2005),
  p.~1267.
\newblock doi: 10.1103/RevModPhys.76.1267.

\bibitem{spengler2012composite}
{\sc C.~Spengler, M.~Huber, and B.~C. Hiesmayr}, {\em Composite
  parameterization and {H}aar measure for all unitary and special unitary
  groups}, Journal of mathematical physics, 53 (2012), p.~013501.
\newblock doi : 10.1063/1.3672064.

\bibitem{wyner1967random}
{\sc A.~D. Wyner}, {\em Random packings and coverings of the unit n-sphere},
  The Bell System Technical Journal, 46 (1967), pp.~2111--2118.

\bibitem{zeh1970interpretation}
{\sc H.~D. Zeh}, {\em On the interpretation of measurement in quantum theory},
  Foundations of Physics, 1 (1970), pp.~69--76.
\newblock doi : 10.1007/BF00708656.

\bibitem{zhang2017spherical}
{\sc K.~Zhang}, {\em Spherical cap packing asymptotics and rank-extreme
  detection}, IEEE Transactions on Information Theory, 63 (2017),
  pp.~4572--4584.

\bibitem{zurek1981pointer}
{\sc W.~H. Zurek}, {\em Pointer basis of quantum apparatus: Into what mixture
  does the wave packet collapse?}, Physical review D, 24 (1981), p.~1516.
\newblock doi : 10.1103/PhysRevD.24.1516.

\bibitem{zurek1982environment}
{\sc W.~H. Zurek}, {\em Environment-induced superselection rules}, Physical
  review D, 26 (1982), p.~1862.
\newblock doi : 10.1103/PhysRevD.26.1862.

\bibitem{zurek2000decoherence}
{\sc W.~H. Zurek}, {\em Decoherence and einselection: The rough guide}, in
  Decoherence: Theoretical, Experimental, and Conceptual Problems: Proceeding
  of a Workshop Held at Bielefeld, Germany, 10--14 November 1998, Springer,
  2000, pp.~309--341.

\bibitem{zurek2003decoherence}
{\sc W.~H. Zurek}, {\em Decoherence, einselection, and the quantum origins of
  the classical}, Reviews of modern physics, 75 (2003), p.~715.
\newblock doi : 10.1103/RevModPhys.75.715.

\end{thebibliography}
\end{document}